# A Beginner's Guide to Bloch Equation Simulations of Magnetic Resonance Imaging Sequences

ML Lauzon, PhD
Depts of Radiology and Clinical Neurosciences, University of Calgary, Calgary, AB, Canada
Seaman Family MR Research Centre, Calgary, AB, Canada
mllauzon@ucalgary.ca

## ABSTRACT

Nuclear magnetic resonance (NMR) concepts are rooted in quantum mechanics, but MR imaging principles are well described and more easily grasped using classical ideas and formalisms such as Larmor precession and the phenomenological Bloch equations. Many textbooks provide in-depth descriptions and derivations of the various concepts. Still, carrying out numerical Bloch equation simulations of the signal evolution can oftentimes supplement and enrich one's understanding. And though it may appear intimidating at first, performing these simulations is within the realm of every imager. The primary objective herein is to provide novice MR users with the necessary and basic conceptual, algorithmic and computational tools to confidently write their own simulator. A brief background of the idealized MR imaging process, its concepts and the pulse sequence diagram are first provided. Thereafter, two regimes of Bloch equation simulations are presented, the first which has no radio frequency (RF) pulses, and the second in which RF pulses are applied. For the first regime, analytical solutions are given, whereas for the second regime, an overview of the computationally efficient, but often overlooked, Rodrigues' rotation formula is given. Lastly, various simulation conditions of interest and example code snippets are given and discussed to help demonstrate how straightforward and easy performing MR simulations can be.

## INTRODUCTION

Magnetic resonance (MR) imaging is one of the most versatile imaging modalities used clinically. It offers a multitude of soft tissue contrasts (often called weightings) such as proton density, $T_1$, $T_2$, susceptibility [1, 2], diffusion [3] and perfusion [4]; it provides various methods of angiography such as time-of-flight [5], phase contrast [6] and contrast-enhanced acquisitions; it allows two- and three-dimensional imaging in any orientation; acquisition in the Fourier domain (*i.e.*, **k**-space) opens it up to advanced reconstruction and image processing techniques (*e.g.*, parallel imaging [7] and compressed sensing [8]); and MR imaging continues to improve and progress.

Although the underlying principles of nuclear magnetic resonance (NMR) are rooted in quantum mechanics, the concepts of MR imaging, and more specifically those of spin-½ particles like hydrogen, are well described and readily understood using classical ideas and equations such as torque (via Larmor precession) and coupled differential equations (via the phenomenological Bloch equations [9]).

There are many excellent textbooks that provide in-depth descriptions and derivations of both the quantum and classical perspectives of NMR, along with the various aspects of MR imaging and spectroscopy. Although we cannot mention them all, a few recommended and suggested texts include those of Slichter [10], Fukushima [11], Liang [12], Brown [13], and de Graaf [14].

To more fully understand and internalize MR imaging concepts, learning from textbooks should ideally be combined with a hands-on approach, *i.e.*, by performing experiments and/or simulations. Carrying out numerical simulations of the MR signal evolution within portions of an imaging sequence can also be helpful and beneficial to optimize acquisition protocols, determine the source of image artifacts, or even develop new acquisition methods.

There are freely available MR simulators, and although we have no affiliation with any of the following nor do we make any claims as to their reliability or usefulness, the interested reader may wish to look at [15, 16, 17] for insight or to get started. That being said, a more fulfilling and personally satisfying stage in one's learning and understanding is that of self-reliance and self-discovery. So, our primary objective herein is to provide novice MR users with the basic conceptual, algorithmic and computational tools to confidently write their own simulator.

We first provide a brief background section of the idealized MR imaging process and concepts, including Larmor precession, the phenomenological Bloch equations and the pulse sequence diagram. In the Theory section, we show that there are effectively two regimes when performing Bloch equation simulations: the first is when only external $z$-directional magnetic fields are present (*i.e.*, no radio frequency [RF] pulses) which has analytical solutions; the second regime includes applied RF pulses, whereby iterative solutions are required. We present an overview of the computationally efficient (but often overlooked) Rodrigues' rotation formula [18] for this regime, an alternative to ordinary differential equation (ODE) solvers [19, 20]. The Methods section gives example code snippets, along with various possible simulation conditions of interest. To enhance readability and facilitate understanding, mathematical derivations have been put into appendices whenever possible.

## BACKGROUND

In MR imaging, the overwhelming majority of clinical and medical research acquisitions are focused on hydrogen atoms whose nucleus, the proton, is a spin-½ particle that has a magnetic dipole moment, $\boldsymbol{\mu}$. When subjected to an external magnetic field, **B**, a dipole moment experiences a torque according to $\boldsymbol{\mu} \times \boldsymbol{B}$ (known as Larmor precession) and precesses at angular frequency $\omega = \gamma|\boldsymbol{B}|$, where $\gamma$ is a proportionality constant called the gyromagnetic ratio. The actual signal measured in an MR imaging sample is called the net magnetization, **M**, and is obtained by summing up all of the individual dipole moments in that sample.

The temporal evolution of this observed *macroscopic* magnetization vector was first proposed in 1946 by Felix Bloch [9]; it includes Larmor precession along with relaxation effects in the Cartesian directions. More specifically, the phenomenological Bloch equations are given by

$$\frac{d\boldsymbol{M}}{dt} = \gamma(\boldsymbol{M} \times \boldsymbol{B}) - \frac{M_x}{T_2}\hat{\boldsymbol{x}} - \frac{M_y}{T_2}\hat{\boldsymbol{y}} + \frac{(M_0 - M_z)}{T_1}\hat{\boldsymbol{z}} \tag{1}$$

where $\boldsymbol{M} = (M_x, M_y, M_z)$, $\boldsymbol{B} = (B_x, B_y, B_z)$, $T_1$ and $T_2$ are the longitudinal and transverse relaxation times, respectively, and $M_0$ is the equilibrium magnetization. Briefly stated, Bloch postulated that the individual dipole moments that comprise **M** within a sample interact with their surrounding environment and with each other. He called these interactions relaxation processes and assumed first-order kinetics such that along the $z$-axis there is regrowth (with time constant $T_1$), whereas in the transverse $xy$-plane there is signal decay (with time constant $T_2$). In biological tissues, this classical (*i.e.*, non-quantum) description has turned out to be suitable and convenient.

The magnetization **M** is the signal of interest, whereas the applied magnetic fields **B** and the relaxation processes determine how **M** varies and evolves over time. In MR imaging, a common and concise manner to represent the time-varying applied magnetic fields is via the pulse sequence diagram, also called the timing diagram (see Figure 1 for an example). The amplitude-modulated (and possibly phase-modulated) RF pulse provides the $B_x$ and $B_y$ components of the **B**-field; and if the RF excitation pulse is frequency-modulated, then it also contributes a $B_z$ component. The main magnetic field, $B_0$, is implied, does not appear in the sequence diagram, is assumed to be perfectly homogeneous and is oriented along the $z$-direction. Lastly, the three idealized magnetic field gradients, $G_x$, $G_y$ and $G_z$ are assumed to be linear changes of $B_z$ with respect to location, *i.e.*, $G_x = \frac{\Delta B_z}{\Delta x}$, $G_y = \frac{\Delta B_z}{\Delta y}$ and $G_z = \frac{\Delta B_z}{\Delta z}$; these magnetic fields are strictly in the $z$-direction and modify $B_z$ in a predictable way to provide spatial localization. Although the sequence diagram and applied fields are assumed to be ideal and perfect both spatially and temporally, inhomogeneities and non-idealities do occur; these can be incorporated as perturbations.

Regardless, one important aspect of the pulse sequence diagram that needs mentioning is its discretized (as opposed to analog or continuous) nature. As Figure 1 shows, the RF and gradient waveforms are pulsed intermittently, sometimes simultaneously or at different times. What is not apparent is that these waveforms are digitized at a temporal resolution, say $\Delta t$, of a few microseconds (typically 1-4 μs) whereby the magnetic fields are presumed to be constant during this sampling time. We make use of this important assumption when we perform actual numerical simulations.

## THEORY

In this section, we provide the mathematical formalisms for two broad regimes, the first where only $B_z$ fields are applied, and the second regime where $B_x$ and/or $B_y$ fields are also present.

### Regime I. Only $B_z$ Fields

When the **B**-field is only along the $z$-direction, the Bloch equations in Eq-1 yield well known analytic solutions. As these are extensively derived in MR textbooks, here we only highlight the salient steps. First, the cross product is expanded into its Cartesian components to obtain three *coupled* ODEs. Next, define the transverse magnetization as $M_{xy} = M_x + iM_y$ with $i = \sqrt{-1}$. Then, set $B_x$ and $B_y$ to zero to obtain two *uncoupled* ODEs in $M_{xy}$ and $M_z$ whose spatially and temporally explicit solutions are

$$\begin{aligned} M_{xy}(\boldsymbol{r},t) &= M_{xy}^0(\boldsymbol{r})\, e^{-t/T_2(\boldsymbol{r})} e^{-i\gamma \int_0^t B_z(\boldsymbol{r},t)dt} \\ M_z(\boldsymbol{r},t) &= M_z^0(\boldsymbol{r})\, e^{-t/T_1(\boldsymbol{r})} + M_0(\boldsymbol{r})\left[1 - e^{-t/T_1(\boldsymbol{r})}\right] \end{aligned} \tag{2}$$

where $M_{xy}^0(\boldsymbol{r})$ and $M_z^0(\boldsymbol{r})$ are initial conditions, parameters $M_0(\mathbf{r})$, $T_1(\mathbf{r})$, and $T_2(\mathbf{r})$ are all tissue specific, $\mathbf{r} = (x,y,z)$, and

$$B_z(\boldsymbol{r},t) = B_0 + \Delta B_z(\boldsymbol{r},t) + \boldsymbol{G}(t)\cdot\boldsymbol{r}(t) \tag{3}$$

The main magnetic field, $B_0$, in Eq-3 produces rotation term $e^{-i\gamma B_0 t}$ in $M_{xy}$ of Eq-2, which effectively represents our resonance condition. In MR imaging, this term is demodulated; internally, pre-acquisition calibration steps determine the center frequency, $\gamma B_0$, after which the measured signal $M_{xy}$ is multiplied by $e^{+i\gamma B_0 t}$. Consequently, the $B_0$ term is conventionally omitted from $B_z(\mathbf{r},t)$, which essentially means that we are dealing with signal $M_{xy}$ in the rotating (as opposed to laboratory) frame of reference.

The second term, $\Delta B_z$, represents any and all off-resonance effects in the MR sample. These can be due to main field inhomogeneities, tissue susceptibility, chemical shift, $z$-directional eddy currents, and so on.

Generally speaking, off-resonance can be both spatially- and time-varying (*e.g.*, eddy currents), but are more often considered to only depend on position (*e.g.*, $B_0$ inhomogeneities, chemical shift).

The last term involves the dot product of magnetic field gradients **G** and position **r**, where we explicitly denoted the temporal dependence of both vectors. The time-varying nature of the gradients is shown in Figure 1. The time dependence of **r** is only applicable when tissues are moving, *e.g.*, arterial and venous blood, respiration, heart motion, peristalsis, or body movement due to restlessness, hiccups, coughing, *etc*. In general, we can expand **r** into its Taylor series to any order. Limiting ourselves to velocity effects (a common practice in clinical MR imaging), we have

$$B_z(\boldsymbol{r}, t) = \Delta B_z(\boldsymbol{r}, t) + \boldsymbol{G} \cdot \boldsymbol{r_o} + \boldsymbol{G} \cdot \boldsymbol{v_o} t \tag{4}$$

where $\mathbf{r_o}$ and $\mathbf{v_o}$ denote initial position and velocity, respectively. We then substitute this $B_z$ into $M_{xy}$ of Eq-2. Next, recall that our sequence is temporally discretized, and we assume that the **B**-field is constant during this sampling time $\Delta t$. With the help of Appendix A, the analytical solutions of Eq-2 at the end of time sample *n*, namely $t = n\Delta t$, are

$$\begin{aligned} M_{xy,n}(\boldsymbol{r}) &= M_{xy}^0(\boldsymbol{r})\, e^{-n\Delta t/T_2(\boldsymbol{r})} e^{-i\gamma[\delta_n(\boldsymbol{r}) + \boldsymbol{r_o}\cdot\boldsymbol{m}_{\boldsymbol{0},n} + \boldsymbol{v_o}\cdot\boldsymbol{m}_{\boldsymbol{1},n}]} \\ M_{z,n}(\boldsymbol{r}) &= M_z^0(\boldsymbol{r}) e^{-n\Delta t/T_1(\boldsymbol{r})} + M_0(\boldsymbol{r})\left[1 - e^{-n\Delta t/T_1(\boldsymbol{r})}\right] \end{aligned} \tag{5}$$

where $\delta_n$ is the cumulative off-resonance term up to end-sample *n*, while $\mathbf{m}_{\mathbf{0},n}$ and $\mathbf{m}_{\mathbf{1},n}$ represent the cumulative zeroth and first moments of gradient $\boldsymbol{G}$. More explicitly,

$$\delta_n(\boldsymbol{r}) = \sum_{k=1}^{n} \Delta B_{z,k}(\boldsymbol{r})\Delta t, \qquad \boldsymbol{m}_{\boldsymbol{0},n} = \sum_{k=1}^{n} \boldsymbol{G}_k \Delta t, \qquad \boldsymbol{m}_{\boldsymbol{1},n} = \sum_{k=1}^{n} \boldsymbol{G}_k \left(k - \tfrac{1}{2}\right) \Delta t^2 \tag{6}$$

**Regime II. Applied $B_x$ and/or $B_y$ Fields**

When the applied **B**-field has components in the *x*- and/or *y*-directions, such as when an RF pulse is applied, Eq-1 yields a set of 3 *coupled* ODEs. Unlike regime-I, ODE coupling means that analytical solutions are no longer viable, so we must resort to solving **M** iteratively.

Referring back to Eq-1, we effectively have two types of terms: (1) Larmor precession via the cross product, and (2) relaxation terms. The cross product is what leads to ODE coupling, whereas relaxation effects are expressed along each Cartesian component individually (*i.e.*, there is no mixing). To solve for **M**, we will first ignore the relaxation terms and present a computationally efficient algorithm, and then modify the algorithm by incorporating relaxation effects.

Rearranging Eq-1 without the relaxation terms yields $d\boldsymbol{M} = \gamma(\boldsymbol{M} \times \boldsymbol{B})dt$. Clearly, d**M** is the *change* in magnetization and is perpendicular to both **M** and **B**. Next, multiply and divide by |**B**|, the magnitude of **B**, recognize $\widehat{\boldsymbol{b}} = \boldsymbol{B}/|\boldsymbol{B}|$ as the unit-vector along **B**, and write $\boldsymbol{M} = \boldsymbol{M}_{\parallel} + \boldsymbol{M}_{\perp}$ where $\boldsymbol{M}_{\parallel}$ and $\boldsymbol{M}_{\perp}$ are the components of **M** that are parallel and perpendicular to $\widehat{\boldsymbol{b}}$, respectively, so that

$$d\boldsymbol{M} = \left(\boldsymbol{M}_{\parallel} \times \widehat{\boldsymbol{b}} + \boldsymbol{M}_{\perp} \times \widehat{\boldsymbol{b}}\right) \gamma |\boldsymbol{B}| dt = \left(\boldsymbol{M}_{\perp} \times \widehat{\boldsymbol{b}}\right) d\theta \tag{7}$$

because $\boldsymbol{M}_{\parallel} \times \hat{\boldsymbol{b}} = 0$ and $d\theta = \gamma|\boldsymbol{B}|dt$. More specifically, this explicitly shows that the (infinitesimal) change in magnetization is the rotation (by angle dθ) of **M** perpendicular to **B**.

Recall that MR sequences are digitized with sample-time Δt on the order of a few microseconds, and that during this sampling time, both the RF and gradients **G** are assumed to be constant (and by extension, so too are $\hat{\boldsymbol{b}}$ and |**B**|). This means that during sampling time Δt, **M** is rotated about $\hat{\boldsymbol{b}}$ (the axis of rotation) by angle $\theta = \gamma|\boldsymbol{B}|\Delta t$. This 3D rotation can be efficiently computed using Rodrigues' rotation formula [18]. As stated previously, one must solve for **M** iteratively at each time sample, whereby the (output) result at a given sample becomes the input for the next sample. The derivation and details of Rodrigues' formula can be found in Appendix B, wherein the magnetization at the end of the $n^{\text{th}}$ sample is given by

$$\boldsymbol{M}_n = \cos\theta_n \boldsymbol{M}_{n-1} + (1-\cos\theta_n)(\hat{\boldsymbol{b}}_n \cdot \boldsymbol{M}_{n-1})\hat{\boldsymbol{b}}_n + \sin\theta_n (\hat{\boldsymbol{b}}_n \times \boldsymbol{M}_{n-1}) \tag{8}$$

Here, $\hat{\boldsymbol{b}}_n = \boldsymbol{B}_n/|\boldsymbol{B}_n|$ is the unit-vector of sample *n*, $\boldsymbol{B}_n = (B_{x,n}, B_{y,n}, B_{z,n})$ are its **B**-field components, $\theta_n = -\gamma|\boldsymbol{B}_n|\Delta t$ is the rotation angle for that sample, and a negative angle is needed because Rodrigues' formula uses the right-hand rule whereas the Bloch equations are defined via the left-hand rule.

Before we include relaxation effects, we first discuss flow (*i.e.*, velocity) effects. Referring back to Eq-4, we see that velocity will induce an explicit time-dependence of $B_z$ *during* sampling time Δt. This clearly violates the assumption that $\mathbf{B}_n$ is constant within sample-time Δt. However, as Δt is small (on the order of 1-4 μs), we *approximate* a constant $B_{z,n}$ at sample *n* by setting $t = \left(n - \frac{1}{2}\right)\Delta t$; this is equivalent to using the $B_{z,n}$ value at the mid-point of the sample, which for velocity-only effects is its *average* value.

Next, let's incorporate relaxation effects. In regime-I the situation was straightforward as **M** rotates strictly perpendicular to $B_z$; this led to separate longitudinal and transverse processes. For regime-II, however, **M** rotates such that both its longitudinal and transverse components are constantly changing. More specifically, |**M**| is non-constant during its rotation θ throughout sample-time Δt. This contravenes Rodrigues' rotation formalism whereby only vector *rotation* is presumed to occur.

We remedy this discrepancy by recognizing that typical biological $T_2$ values are on the order of tens of milliseconds, while their corresponding $T_1$ values are 3-20 times greater than $T_2$. And since Δt is on the order of only a few microseconds (*i.e.*, $10^{-4}$ to $10^{-6}$ times smaller than relaxation times), we approximate the relaxation effect at each end-sample time via

$$\boldsymbol{M}_n = \begin{bmatrix} M_{x,n} \\ M_{y,n} \\ M_{z,n} \end{bmatrix} \rightarrow \begin{bmatrix} M_{x,n}\left(1-\frac{\Delta t}{T_2}\right) \\ M_{y,n}\left(1-\frac{\Delta t}{T_2}\right) \\ M_{z,n}\left(1-\frac{\Delta t}{T_1}\right) + M_0\frac{\Delta t}{T_1} \end{bmatrix} \tag{9}$$

Effectively, we have a two-step process at each time sample: (1) calculate $\mathbf{M}_n$ using Eq-8, and (2) if relaxation effects are to be included, modify $\mathbf{M}_n$ of step-1 using Eq-9.

## METHODS

Next, we turn our attention to actually performing numerical Bloch equation simulations. The Theory section provided the road map, so we now need to translate this to usable algorithms. And because our goal is to guide users into creating and exploring Bloch simulations under various conditions on their own, we provide a few representative simulation examples along with their respective code snippets.

There are many available programming languages, be they compiled (such as C or C++) or interpreted (*e.g.*, Python and Matlab). And although computer algorithms are often shown as pseudocode, here we present the code snippets using Matlab (The MathWorks Inc., Natick, MA). The reasons are that Matlab is used pervasively in the scientific, academic and MR communities; as an interpreted language it is easy to use (even for novices); it comprises an extensive set of available numerical and graphical functions; it has considerable help documentation; and it lends itself to short, easy-to-read and easy-to-write code.

Within the presented code snippets, vectors and arrays are indexed from 1 (as opposed to 0), variable broadcasting is automatic, and using "vectorization" to avoid for-loops is the preferred method. The sequence components (RF and/or gradients) are presumed to exist as vectors, sampling-time $\Delta t$ is known, whereas the voxel and/or tissue characteristics and the calculation steps are given explicitly.

Lastly, the examples have been inspired from actual vendor MR imaging sequences, in this case General Electric Healthcare (Chicago, USA). As such, the code snippets herein follow their unit conventions, namely the SI-cgs system of units with RF amplitude in gauss (G), gradient amplitude in G/cm, and velocity in cm/s. Other vendors may use the SI-mks system of units whereby RF and gradient amplitudes are in μT and mT/m, respectively. Regardless of the unit-system used, *consistency* of units is imperative, and conversions may be necessary (*e.g.*, T is $10^4$ G, mT/m is 0.1 G/cm, Hz is $2\pi$ rad/s).

**Simulation 1**. Movement and flow, especially from fast-flowing blood, can lead to significant ghosting artifacts in the resulting MR image. To mitigate this effect, one often applies a gradient-moment nulling technique (*aka*, flow compensation) so that the *net* phase accrual of flowing spins is zero. We wish to verify that this is indeed the case. Let's assume that an appropriate tripolar $G_x$ waveform on the readout axis has been defined, and that $M_{xy}^0(\boldsymbol{r})$ is the known initial transverse magnetization. This imaging sequence portion only involves $G_x$, so we are clearly in regime-I. We can determine the magnetization evolution and phase accrual over time for a single position and velocity (Code Snippet 1). Conversely, we can extend this example to simultaneously explore the net phase accrual at some specific time (*e.g.*, the echo time TE) for a range of positions and velocities (Code Snippet 2).

The subsequent simulations are for regime-II, and all make use of Eq-8. The corresponding algorithm to calculate the unit-vector $\hat{\boldsymbol{b}}$ is shown in Code Snippet 3, whereas the actual implementation of Rodrigues' rotation formula is given in Code Snippet 4.

**Simulation 2**. For our first simulation in regime-II, we calculate the magnetization of a non-slice-selective, amplitude-modulated, 90° excitation RF pulse. We assume that the pulse is oriented along the *x*-axis and that the initial magnetization is strictly along the *z*-axis (see Code Snippet 5). By evaluating over a range of frequencies, one can observe and characterize the slice profile and full width at half maximum (FWHM). One can readily extend this example to simultaneously look at the effect of varying the flip angle (as shown in Code Snippet 6). In this case, we calculate the magnetization as a function of both frequency and a $B_1$ sensitivity factor $\leq 1$ (this acts to reduce the desired flip angle). One would observe that the slice profile shapes, transition zones and FWHM vary as a function of flip angle. We can also readily extend these examples to incorporate relaxation effects (Code Snippet 7), be slice-selective, or include flow effects (see Code Snippet 8).

**Simulation 3**. The next examples in regime-II evaluate AM/PM and AM/FM radio frequency pulses. More specifically, we could look at an adiabatic fast passage 180º inversion pulse, as suggested by the seminal work of Silver [21], as a function of frequency. In its AM/PM representation, we have RF and $RF_{phs}$ axes / waveforms in the sequence diagram, as explained in the caption of Figure 1. These two waveforms can be combined into an effective complex waveform, after which $B_x$ and $B_y$ are the real and imaginary parts, respectively. So, looking at Code Snippet 5, we immediately see that it is also applicable to AM/PM radio frequency pulses, provided that (code-snippet variable) "vecRF" is indeed the complex representation of the RF pulse. If, however, the RF pulse is given by its AM/FM representation, the situation is different as we now have an $RF_{frq}$ axis (and waveform). In this case, a slight modification of the source code is necessary, as shown in Code Snippet 9. As before, the RF pulse could be made slice-selective with the use of gradients, and relaxation and/or flow effects could also be included.

**Simulation 4**. Our last illustration in regime-II is the quick-and-easy design and evaluation of a two-dimensional RF pulse based on the principles of excitation **k**-space [22]. Knowledge of the $G_y$ and $G_z$ waveforms allows one to calculate the desired $(k_y,k_z)$-trajectory, while numerical simulation of the RF pulse (see Code Snippet 10) lets one characterize the effective 2D slice profile.

The examples presented herein, along with the somewhat abridged code snippets, demonstrate the ease of simulating single or multi-dimensional RF pulses, be they AM, PM, FM, and also show how these can be readily evaluated for various voxel conditions.

## DISCUSSION

We briefly presented a few important concepts of MR imaging, namely Larmor procession, the phenomenological Bloch equations, relaxation processes and their time constants ($T_1$ and $T_2$), and how they relate to the evolution of magnetization. There are, of course, many other important MR concepts, such as the rotating wave approximation, J-coupling [23], the extended phase graph [24], or the spin density matrix, all of which are detailed in MR textbooks (see suggested list in the Introduction). These advanced topics help explain and predict the observed and evolving magnetization under specific and/or various conditions. Here, however, we did not include these more advanced effects as our goal was geared towards establishing the fundamental basis of a numerical Bloch equation simulation.

That being said, some advanced concepts can be understood or explored using the methods presented herein. For example, the extended phase graph concept is the generalization of stimulated echoes in the presence of many (*i.e.*, 4 or more) RF pulses. Briefly stated, a stimulated echo results whenever three RF pulses are played out "close in time" (*i.e.*, close with respect to the relaxation processes); actual RF pulse profiles produce magnetization of different proportions along the longitudinal axis ($\pm M_z$) and in the transverse planes ($M_{xy}$), even for so-called 90º or 180º pulses; for 2 RF pulses, only a primary echo is formed, but when a 3$^{rd}$ RF pulse is applied, both a primary *and* a secondary (*i.e.*, stimulated) echo appear, all of which can be verified (and perhaps better understood) performing numerical Bloch simulations. One can then extend the simulation to 4 (or more) RF pulses. Conversely, one could simulate various multi-RF sequences with different timings and/or flip angles and compare the end results. In either case, one gains understanding, be it conceptual or practical, of an advanced MR topic or notion.

Magnetic resonance imaging is conventionally taught and explained assuming ideal cases, such as a perfectly homogeneous $B_0$ field, truly linear gradients whose magnetic fields are strictly in the *z*-direction, digitized waveforms with unchanging RFs and gradients during sampling $\Delta t$, no $B_1$ inhomogeneity, no noise, and so on. We followed this common didactic practice, but realistically these conditions are never met. However, one can include and simulate some of these non-idealities. For example, $B_0$ inhomogeneity can be explored by simulating voxels at different $\Delta f_z$ frequencies, $B_1$ inhomogeneity can be explored by including an RF factor (as done in Code Snippet 6), effects of time-varying concomitant magnetic fields and/or eddy currents in the *x*- or *y*-directions can be added (note that this puts us in regime-II), *etc*. In effect, non-idealities are modeled as *perturbations* to the ideal experiment. Although it may not always be exact or fully representative of the real-world MR acquisition, one should recognize that numerical simulations are often used *a priori* to build, design or pre-optimize an imaging sequence, or performed *a posteriori* to gain some insight, optimize the sequence, or explain observed artifacts.

When it comes to numerical simulations, the overwhelming majority of the MR community uses ODE solvers, with very little mention of Rodrigues' formula. The coupled ODE nature of the Bloch equations naturally leads one to using ODE solvers. But, as shown here, it is not the *only* viable algorithm. It may be that Rodrigues' rotation formula is not well known, or perhaps it is perceived inferior to ODE solvers. To be fair, ODE solvers are more generalizable, can accommodate extra terms (such as those needed for relaxation and/or diffusion processes [25]), and when written in a compiled language, the executables are computationally fast and efficient.

ODE solvers, by their very nature, are iterative *within* sampling-time $\Delta t$. This means that they calculate intermediary values at multiple sub-sampling times to find the result at end-time $\Delta t$. For compiled languages, this is not an issue. However, user-friendly interpreted languages like Matlab benefit from vectorized (*i.e.*, non-for-loop) algorithmic implementations like Rodrigues' rotation formula. Moreover, Code Snippet 4 shows that it can be readily coded (and hopefully understood) in only a few lines. So, although regime-II requires one to iteratively solve **M** at each time-sample $\Delta t$, Rodrigues' formula eliminates the *internal* iteration that ODE solvers perform to get the end-sample result. This is what makes Rodrigues' formula so much more computationally efficient, especially in Matlab.

And, from an academic perspective, a 3D rotation of one vector about a unit-vector is conceptually easier to grasp than the numeric solution of coupled ODEs. The hope is that this simple and overarching MR concept helps to better explain and/or understand the direct effects of magnetic fields on the observed and evolving magnetization signal.

In the ideal case where only $B_z$ magnetic fields are applied, we showed that analytical solutions exist, so that iterative ODE solvers or Rodrigues' rotation formula are not required for this regime. Although this "ideal" assumption is, for all intents and purposes unrealistic, it still provides an informative, albeit simplistic, description of the magnetization evolution. Similarly, in regime-II where RF pulses are applied, Rodrigues' rotation formula (and possibly modifying it to include relaxation effects) describes the magnetization, provided that the **B**-field within sampling-time $\Delta t$ remains effectively constant. Again, this "constant **B**" condition is most probably not strictly met in reality, but for sampling times on the order of a few microseconds, this may not be of great concern.

Although performing numerical Bloch equation simulations may yield inexact solutions (with respect to reality), they are still typically good-to-excellent approximations to reality, and they provide concrete and valuable information that can enhance one's understanding. So, to help novice MR researchers write and interact with their own simulator, and to help guide them along the way, we presented the conceptual and algorithmic basics along with a few representative examples with code snippets.

## ACKNOWLEDGMENTS

The authors acknowledge funding from the Canadian Institutes of Health Research (CIHR), the Heart and Stroke Foundation of Canada (HSFC), and the Hopewell Professorship.

# APPENDICES

## A. Integration of $B_z$ in Regime I

First, the time integral of $\Delta B_z(\mathbf{r},t)$. Within sample-time $\Delta t$, we assume that $\Delta B_z$ is temporally constant; in other words, for time sample *k* we have $\Delta B_{z,k}(\mathbf{r})$. Thus, at the *end* of time sample *n* we have

$$\begin{aligned}\int_0^{n\Delta t} \Delta B_z(\boldsymbol{r},t)dt &= \int_0^{\Delta t} \Delta B_{z,1}(\boldsymbol{r})dt + \int_{\Delta t}^{2\Delta t} \Delta B_{z,2}(\boldsymbol{r})dt + \cdots + \int_{(n-1)\Delta t}^{n\Delta t} \Delta B_{z,n}(\boldsymbol{r})dt \\ &= \sum_{k=1}^{n} \Delta B_{z,k}(\boldsymbol{r})\,\Delta t\end{aligned} \tag{10}$$

Second, we deal with the gradient terms in Eq-4. The p[th] moment of gradient $\mathbf{G}$ is defined as $\int_0^t \boldsymbol{G}t^p dt$. The gradients are discretized with sample-time $\Delta t$ and assumed constant within that duration so that at time sample *k* the gradient amplitude is $\mathbf{G}_k$. Thus, the p[th] gradient moment up to end-sample *n* is

$$\begin{aligned}\int_0^{n\Delta t} \boldsymbol{G}t^p dt &= \int_0^{\Delta t} \boldsymbol{G}_1 t^p dt + \int_{\Delta t}^{2\Delta t} \boldsymbol{G}_2 t^p dt + \cdots + \int_{(n-1)\Delta t}^{n\Delta t} \boldsymbol{G}_n t^p dt \\ &= \sum_{k=1}^{n} \boldsymbol{G}_k \int_{(k-1)\Delta t}^{k\Delta t} t^p dt = \sum_{k=1}^{n} \boldsymbol{G}_k \frac{[k^{p+1} - (k-1)^{p+1}]}{p+1} \Delta t^{p+1}\end{aligned} \tag{11}$$

More specifically, the first few moments for $p = \{0,1\}$ are, respectively,

$$\sum_{k=1}^{n} \boldsymbol{G}_k \Delta t, \qquad \sum_{k=1}^{n} \boldsymbol{G}_k \left(k - \tfrac{1}{2}\right) \Delta t^2 \tag{12}$$

## B. Rodrigues' Rotation Formula in Regime II

Larmor precession is proportional to $\boldsymbol{M} \times \boldsymbol{B}$, which we showed leads to $\boldsymbol{M} \times \hat{\boldsymbol{b}}$. This follows the *left*-hand rule. By convention, however, Rodrigues' rotation formula is the rotation of vector $\mathbf{V}$ about unit-vector $\hat{\boldsymbol{u}}$ by angle $\theta$ and follows the *right*-hand rule, namely $\boldsymbol{V}_{\text{rot}} = \hat{\boldsymbol{u}} \times \boldsymbol{V}$. This difference in handedness is easily accommodated by negating the angle of rotation.

Next, we derive Rodrigues' formula. Let $\boldsymbol{V} = \boldsymbol{V}_{\parallel} + \boldsymbol{V}_{\perp}$, where $\boldsymbol{V}_{\parallel}$ and $\boldsymbol{V}_{\perp}$ are the components of $\boldsymbol{V}$ that are parallel and perpendicular to $\hat{\boldsymbol{u}}$, respectively, so $\boldsymbol{V}_{\text{rot}} = \boldsymbol{V}_{\parallel,\text{rot}} + \boldsymbol{V}_{\perp,\text{rot}}$. Clearly $\boldsymbol{V}_{\parallel,\text{rot}} = \boldsymbol{V}_{\parallel}$. Since $\boldsymbol{V}_{\perp}$ and $\hat{\boldsymbol{u}}$ are orthogonal to one another, then $\hat{\boldsymbol{u}} \times \boldsymbol{V}_{\perp}$ is orthogonal to both $\boldsymbol{V}_{\perp}$ and $\hat{\boldsymbol{u}}$. So, the rotation of $\boldsymbol{V}_{\perp}$ by $\theta$ about $\hat{\boldsymbol{u}}$ is just the vector sum of the rotated components along $\boldsymbol{V}_{\perp}$ and along $\hat{\boldsymbol{u}} \times \boldsymbol{V}_{\perp}$. Putting all of these together, we obtain

$$\boldsymbol{V}_{\text{rot}} = \boldsymbol{V}_{\parallel} + \cos\theta\, \boldsymbol{V}_{\perp} + \sin\theta\, (\hat{\boldsymbol{u}} \times \boldsymbol{V}_{\perp}) \tag{13}$$

Next, we (a) note that $\hat{\boldsymbol{u}} \times \boldsymbol{V} = \hat{\boldsymbol{u}} \times (\boldsymbol{V}_\parallel + \boldsymbol{V}_\perp) = \hat{\boldsymbol{u}} \times \boldsymbol{V}_\perp$, (b) substitute $\boldsymbol{V}_\perp = \boldsymbol{V} - \boldsymbol{V}_\parallel$ into the second term above, and (c) use $\boldsymbol{V}_\parallel = (\hat{\boldsymbol{u}} \cdot \boldsymbol{V})\hat{\boldsymbol{u}}$ to finally arrive at Rodrigues' rotation formula, namely

$$\boldsymbol{V}_{\mathrm{rot}} = \cos\theta\, \boldsymbol{V} + (1 - \cos\theta)(\hat{\boldsymbol{u}} \cdot \boldsymbol{V})\hat{\boldsymbol{u}} + \sin\theta\, (\hat{\boldsymbol{u}} \times \boldsymbol{V}) \tag{14}$$

More explicitly, in MR imaging vector **V** becomes magnetization **M**, while the rotation axis $\hat{\boldsymbol{u}}$ becomes unit-vector $\hat{\boldsymbol{b}}$. And because we *iteratively* need to step through the time samples, the input to time sample *n* is $\mathbf{M}_{n-1}$, whereas the resultant (*i.e.*, rotated) magnetization is $\mathbf{M}_n$.

FIGURES

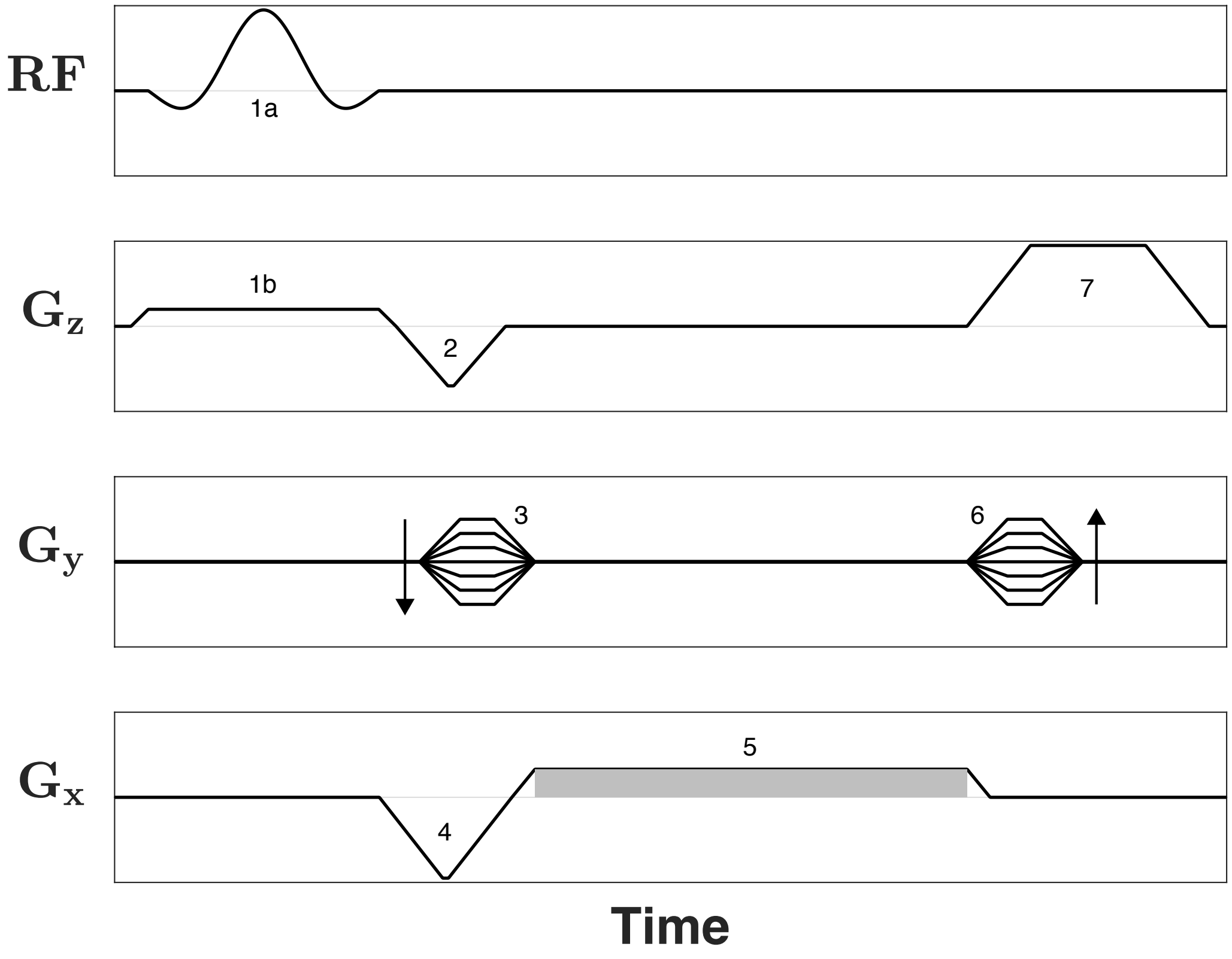

Figure 1: Pulse sequence diagram of the spoiled gradient-recalled echo sequence. The horizontal axis is time, whose units are typically given in ms. All three gradient waveforms are shown, where $G_x = \frac{\Delta B_z}{\Delta x}$, $G_y = \frac{\Delta B_z}{\Delta y}$ and $G_z = \frac{\Delta B_z}{\Delta z}$, along with the amplitude-modulation (AM) portion of the RF waveform. If the excitation pulse is also phase-modulated (PM) and/or frequency-modulated (FM), then supplemental $RF_{phs}$ and/or $RF_{frq}$ axes are required. The units along the axes are usually omitted, but RF amplitudes are often stated in μT (or G), those of $RF_{phs}$ and $RF_{frq}$ are in rad and Hz, respectively, while gradients are given in mT/m (or G/cm). The paired labelled pulses 1a/1b represent slice-selective excitation, whereas pulse 2 along $G_z$ is the slice-select rephaser. On the $G_y$ axis, pulses 3 and 6 are the phase-encode and (opposite polarity) phase-rewinder lobes, respectively; the arrows denote the direction of the gradient amplitude on subsequent excitations to acquire the entire **k**-space dataset. On the $G_x$ axis, pulse 4 is sometimes called the readout prephaser, while pulse 5 is referred to as the readout. Albeit confusing nomenclature, the $G_x$ axis behaves like all other gradients and is not, in and of itself, actively acquiring (or reading) the **k**-data. However, data acquisition (*i.e.*, reading out) is commonly denoted with a shaded area (as done here); an equivalent but less frequent way is to add a DAQ (data acquisition) axis to the pulse sequence diagram and denote signal acquisition via rectangular pulses. Regardless of the scheme used, data acquisition occurs *simultaneously* with $G_x$ pulse 5, hence its moniker as the readout pulse. Lastly, pulse 7 on the $G_z$ axis is called an end-of-sequence gradient killer or spoiler; its purpose is to dephase any remnant $M_x$ and $M_y$ transverse magnetization prior to the subsequent RF excitation.

## CODE SNIPPETS

Code Snippet 1: Flow-compensation (part 1). Calculate the temporal evolution of the transverse magnetization components $M_x$ and $M_y$, along with the phase accrual when subjected to a $G_x$ waveform. Here, we specify a single voxel condition with known position ($x_0$) and velocity ($v_x$), then calculate the magnetization at multiple time points based on the zeroth- and first-order gradient moments.

```
% Calculate various gradient moments
  N = length( vec_Gx );
  k = 1 : N;

  tim  = k * dt;
  m0_x = cumsum( vec_Gx ) * dt;
  m1_x = cumsum( (k - 1/2) .* vec_Gx ) * dt^2;

% Set voxel conditions and initial Mxy
  x0 = 0.02;      % [cm]
  vx = 7.5;       % [cm/s]
  Mxy_0 = exp( 1i * pi/2 );

% Calculate Mxy at all time points of sequence
  GAM = 2 * pi * 4257.59;     % [rad/s/G]
  phi = GAM * (x0 * m0_x + vx * m1_x);
  Mxy = Mxy_0 * exp( -1i * phi );

% Calculate phase accrual wrt Mxy_0
  phs = angle( Mxy / Mxy_0 );
```

Code Snippet 2: Flow compensation (part 2). Calculate the transverse magnetization components ($M_x$ and $M_y$) and the phase accrual at echo time TE when subjected to a $G_x$ waveform. Here, we specify voxels as a 2D matrix of ($x_0$, $v_x$) and calculate the net phase magnetization at echo time TE within the sequence.

```
% Calculate various moments up to time TE (idx_TE)
  k    = 1 : idx_TE;
  m0_x = sum( vec_Gx ) * dt;
  m1_x = sum( (k - 1/2) .* vec_Gx ) * dt^2;

% Set voxel conditions and initial Mxy
%   x0 in [cm], vx in [cm/s]
  [ vx, x0 ] = ndgrid( -50:50, -10:0.1:10 );
  Mxy_0 = exp( 1i * pi/2 );

% Calculate Mxy at all (x0,vx) at time TE
  GAM = 2 * pi * 4257.59;     % [rad/s/G]
  phi = GAM * (x0 * m0_x + vx * m1_x);
  Mxy = Mxy_0 * exp( -1i * phi );

% Calculate phase change wrt Mxy_0
  phs = angle( Mxy / Mxy_0 );
```

Code Snippet 3: Function to calculate $\hat{\boldsymbol{b}}$. The input **B**-field components ($B_x$, $B_y$, $B_z$) are in gauss and the sampling-time ($\Delta t$) in seconds; the output components of $\hat{\boldsymbol{b}}$ are unitless, whereas the angle is in radians.

```
function [bx, by, bz, ang] = calcBhat( Bx, By, Bz, dt )

  % Bx,By,Bz in [G], dt in [s], GAM in [rad/s/G]

  % Set length/magnitude of B
    tol  = 1e-14;
    Blen = sqrt( Bx.^2 + By.^2 + Bz.^2 );
    idx  = ( Blen <= tol );
    Blen( idx ) = tol;

  % Set rotation-axis (unit-vector) components
    bx = Bx ./ Blen;
    by = By ./ Blen;
    bz = Bz ./ Blen;

  % Set rotation-angle [rads], note negative sign
    GAM = 2 * pi * 4257.59;
    ang = -GAM * Blen * dt;
    ang( idx ) = 0;

end
```

Code Snippet 4: Rodrigues' rotation formula. Function to calculate generic vector **V** rotated about generic unit-vector $\hat{\boldsymbol{k}}$. Input theta is in radians.

```
function [Rx, Ry, Rz] = calcRRF( Vx, Vy, Vz, kx, ky, kz, theta )

  % By calculating (x,y,z) components explicitly,
  %   (V,k,theta) can be N-D, but all of same size

  % Calculate cross(kHat,V) explicitly
    crsKV_x = ky .* Vz - kz .* Vy;
    crsKV_y = kz .* Vx - kx .* Vz;
    crsKV_z = kx .* Vy - ky .* Vx;

  % Calculate dot(kHat,V)kHat explicitly
    dotKV   = kx .* Vx + ky .* Vy + kz .* Vz;
    dotKV_x = dotKV .* kx;
    dotKV_y = dotKV .* ky;
    dotKV_z = dotKV .* kz;

  % Calculate trigonometric functions
    c = cos( theta );
    s = sin( theta );

  % Do vectorized multiplication (fast and efficient)
    Rx = (c .* Vx) + (s .* crsKV_x) + (1 - c) .* dotKV_x;
    Ry = (c .* Vy) + (s .* crsKV_y) + (1 - c) .* dotKV_y;
    Rz = (c .* Vz) + (s .* crsKV_z) + (1 - c) .* dotKV_z;

end
```

Code Snippet 5: RF excitation (part 1). Calculate the magnetization and effective slice profile of a non-slice-selective, amplitude-modulated, excitation RF pulse over a range of frequencies ($\Delta f_z$). Here, we assume that relaxation and flow effects can be neglected.

```
% Set constants
  GAM = 4257.59;         % [Hz/G]

% Set simulation conditions
%   Mxy_0=[0;0;1], dfz range in [Hz], no relaxation/flow
  Mx  = 0;  My = 0;  Mz = 1;
  dfz = -800 : 0.1 : 800;
  dBz = dfz / GAM;

% Perform simulation (iterating over RF samples)
%   Note that Bz = dBz
  for cnt = 1 : length(vec_RF)
    Bx = real( vec_RF(cnt) );
    By = imag( vec_RF(cnt) );

    [bx, by, bz, ang] = calcBhat( Bx, By, dBz, dt );
    [Mx, My, Mz] = calcRRF( Mx, My, Mz, bx, by, bz, ang );
  end
```

Code Snippet 6: RF excitation (part 2). Calculate the magnetizations and normalized slice profiles of a non-slice-selective, amplitude-modulated, excitation RF pulse over a range of frequencies ($\Delta f_z$) and flip angles, the latter of which is achieved via a $B_1$ sensitivity factor.

```
% Set constants
  GAM = 4257.59;         % [Hz/G]

% Set simulation conditions
%   Mxy_0=[0;0;1], dfz range in [Hz], no relaxation/flow
  Mx  = 0;  My = 0;  Mz = 1;
  fac = 0.16 : 0.01 : 1;
  dfz = -800 : 1 : 800;

  [ fac, dBz ] = ndgrid( fac, dfz/GAM );

% Perform simulation (iterating over RF samples)
%   Note that Bz = dBz
  for cnt = 1 : length(vec_RF)
    Bx = fac * real( vec_RF(cnt) );
    By = fac * imag( vec_RF(cnt) );

    [bx, by, bz, ang] = calcBhat( Bx, By, dBz, dt );
    [Mx, My, Mz] = calcRRF( Mx, My, Mz, bx, by, bz, ang );
  end
  Mxy = abs( Mx + 1i * My);

% Normalize Mxy profiles to 100% at each fac value
  prf_max = max( Mxy, [], 2 );
  Mxy_nrm = 100 * Mxy ./ prf_max;
```

Code Snippet 7: RF excitation (part 3). Calculate the magnetizations and normalized slice profiles of a non-slice-selective, amplitude-modulated, excitation RF pulse over a range of frequencies ($\Delta f_z$) and T2 values. Note the consistency of units via conversion.

```
% Set constants
  GAM = 4257.59;          % [Hz/G]

% Set simulation conditions
%   Mxy_0=[0;0;1], dfz range in [Hz], need T2 in [s]
  Mx  = 0;  My = 0;  Mz = 1;
  T2  = 10 : 10 : 1000;         % [ms]
  dfz = -800 : 1 : 800;

  [ T2, dBz ] = ndgrid( T2 * ms_to_s, dfz/GAM );

% Pre-calculate T2 relaxation factor
  fac_T2 = 1 – dt ./ T2;

% Perform simulation (iterating over RF samples)
%   Note that Bz = dBz
  for cnt = 1 : length(vec_RF)
    Bx = fac * real( vec_RF(cnt) );
    By = fac * imag( vec_RF(cnt) );

    [bx, by, bz, ang] = calcBhat( Bx, By, dBz, dt );
    [Mx, My, Mz] = calcRRF( Mx, My, Mz, bx, by, bz, ang );
    Mx = Mx .* fac_T2;
    My = My .* fac_T2;
  end
  Mxy = abs( Mx + 1i * My);

% Normalize Mxy profiles to 100% at each fac value
  prf_max = max( Mxy, [], 2 );
  Mxy_nrm = 100 * Mxy ./ prf_max;
```

Code Snippet 8: RF excitation (part 4). Calculate the magnetizations and slice profiles of a slice-selective, amplitude-modulated, excitation RF pulse over a range of positions ($p_z$) and velocities ($v_z$). Note the consistency of units via conversion, and the explicit use of the "average" time when evaluating $B_z$.

```
% Set simulation conditions
%   Mxy_0=[0;0;1], need posZ in [cm], velZ in [cm/s]
  Mx = 0;  My = 0;  Mz = 1;
  pz = -3 : 6 / 400 : 3;       % [mm]
  vz = -100 : 0.5 : 100;       % [cm/s]

  mm_to_cm = 0.1;
  [ velZ, posZ ] = ndgrid( vz, pz * mm_to_cm);

% Perform simulation (iterating over RF samples)
  for cnt = 1 : length(vec_RF)
    Bx = real( vec_RF(cnt) );
    By = imag( vec_RF(cnt) );
    Gz = vec_Gz( cnt );
    Bz = (posZ + (cnt-0.5) * velZ * dt) * Gz;

    [bx, by, bz, ang] = calcBhat( Bx, By, Bz, dt );
    [Mx, My, Mz] = calcRRF( Mx, My, Mz, bx, by, bz, ang );
  end
  Mxy = abs( Mx + 1i * My );
```

Code Snippet 9: Frequency-modulated RF pulse. Calculate the magnetization components of an amplitude- and frequency-modulated (AM/FM) radio frequency pulse over a range of frequencies ($\Delta f_z$). Here, we assume that relaxation and flow effects can be neglected.

```
% Set constants
  GAM = 4257.59;         % [Hz/G]

% Set simulation conditions
%   Mxy_0=[0;0;1], dfz range in [Hz], no relaxation/flow
  Mx  = 0;  My = 0;  Mz = 1;
  dfz = -1200 : 0.1 : 1200;
  dBz = dfz / GAM;

% Perform simulation (iterating over RF samples)
  for cnt = 1 : length(vec_RF)
    Bx  = real( vec_RF(cnt) );
    By  = imag( vec_RF(cnt) );
    B1z = vec_RFfrq(cnt) / GAM;

    Bz  = B1z + dBz;

    [bx, by, bz, ang] = calcBhat( Bx, By, dBz, dt );
    [Mx, My, Mz] = calcRRF( Mx, My, Mz, bx, by, bz, ang );
  end
```

Code Snippet 10: 2D RF excitation. Calculate the magnetization components and effective two-dimensional slice profile of a (*y,z*) slice-selective RF excitation pulse. Note the consistency of units via conversion, and the fact that relaxation and flow effects are omitted.

```
% Set simulation conditions
%   Mxy_0=[0;0;1], need (posY,posZ) in [cm]
  Mx = 0;  My = 0;  Mz = 1;
  y  = -50 : 100 / 300 : 50;    % [mm]
  z  = -3  :  6 / 300  : 3;     % [mm]

  mm_to_cm = 0.1;
  [ posY, posZ ] = ndgrid( y * mm_to_cm, z * mm_to_cm );

% Perform simulation
  for cnt = 1 : length(vec_RF)
    Gy = vec_Gy( cnt );
    Gz = vec_Gz( cnt );
    Bx = real( vec_RF(cnt) );
    By = imag( vec_RF(cnt) );
    Bz = Gy * posY + Gz * posZ;

    [bx, by, bz, ang] = calcBhat( Bx, By, Bz, dt );
    [Mx, My, Mz] = calcRRF( Mx, My, Mz, bx, by, bz, ang );
  end
  Mxy = abs( Mx + 1i * My);
```